\documentclass[preprint, 12pt]{elsarticle}
\usepackage[top=24truemm, bottom=24truemm, left=26truemm, right=26truemm]{geometry}

\usepackage{lineno,hyperref}
\usepackage{amsmath}
\usepackage{amssymb}
\usepackage{amsfonts}

\usepackage[]{graphicx}
\usepackage[dvipdfmx]{}
\usepackage{subfig}
\usepackage{comment}
\usepackage{multicol}
\usepackage{amsmath,amssymb}
\usepackage[font=footnotesize,labelfont=bf]{caption}

\makeatother

\journal{Journal of \LaTeX\ Templates}

\newcommand{\Equref}[1]{Equation \ref{#1}}
\newcommand{\Figref}[1]{Figure \ref{#1}}

\begin{document}
\title{\bf{Simulation study on impact of \\GEM geometry for gas gain uniformity}}

\author{T. Ogawa\footnote{E-mail: ogawat@post.kek.jp}, Y. Aoki \\ \vspace{+2mm} on behalf of the LCTPC-Japan group}
\address{\vspace{+2mm}The Graduate University for Advanced Studies (SOKENDAI), Tsukuba 305-0801, Japan}

\begin{frontmatter} 
\begin{abstract}
Gas Electron Multiplier (GEM) is one of the devices for gas amplification and available as an amplification part of detectors for many experiments. A GEM with a 50 $\mu m$ thick insulator is popular and widely used in many experimental fields. However it is necessary to use several layers of them to get sufficient gas gain because gas gain which is provided with only one 50 $\mu m$ thick GEM is only several times ten. On the other hand, a thick GEM whose amplification area is expanded from 50 to 100 $\mu m$ to get higher gas gain which can be sufficient even with a double stack configuration. But the measurement of gas gain uniformity using a large area 100 $\mu m$ thick GEM reported here observed sizable non-uniformity which reached more than 50\%. We investigated gas gain uniformity which is derived from geometries of the GEM using $\rm{garfield^{++}}$ and considered an optimal geometry for the 100 $\mu m$ thick GEM. 
\end{abstract}
\end{frontmatter} 


\section{Introduction} \label{Section1}
Since F.Sauli or Y.Geomataris developed Micro Pattern Gaseous Detectors (MPGDs) such as GEM \cite{Aus1} or Micro-MEsh GAseous Structure (MicroMEGAS \cite{Aus2}),  MPGDs have been widely used as a part of detectors for radiation detection or imaging in many fields like particle and nuclear physics experiments, astro experiments, and medical imaging. MPGDs will also play an important part of detectors in the current and future most advanced experiments. Particularly in collider experiments, the active area of a GEM or a MicroMEGAS needs to be enlarged in order to cover the most part of a huge detector. Moreover, because mechanical structures which support MPGDs will cause dead spaces, a large area GEM foil or MicroMEGAS mesh is needed.      
%
Concerning the geometry of the GEM, for instance, a insulator is sandwiched by thin conductors like copper and holes which pierce the GEM foil are regularly aligned. By applying a high voltage to the copper electrodes, a high electric field is formed in the holes. Electrons are accelerated near or in the holes and the avalanche amplification process is induced. Because electric potential on the both copper electrodes of the same GEM foil is fixed, in the case that the thickness of the insulator is different in some places, a degree of avalanche amplification will eventually vary since the high electric field which is formed in the GEM holes is also different. This effect causes a degradation of energy resolution and other problems. If more high voltage is applied to recover a part of the poor amplification area, it will cause discharges and damage to the detectors.
 %
%
On the future International Linear Collider (ILC) project, Time Projection Chamber (TPC) is a candidate for a main tracker and the GEMs are also a candidate of an amplification device. Currently, there are two kinds of GEMs which are widely used. One is a so-called standard GEM whose thickness of the insulator is 50 $\mu m$. Another one is a thick GEM whose thickness of the insulator is 100 $\mu m$. The standard GEM is basically used with a multilayered configuration because gas gain which we get from only one standard GEM without discharges is not very high. And also there are disadvantages that high voltage lines and mechanical structures to support GEMs are complicated. In contrast, a double stacked configuration of the 100 $\mu m$ thick GEM provides sufficient gas gain which is approximately $O(10^{4})$ because gas gain we can get from one thick GEM is larger than that of the standard GEM. Since the number of stacked GEMs with the thick GEM configuration is less, handling of mechanics of such configuration will be also simpler than that with the standard GEM.
%
%
\begin{figure}[htbp]
\begin{center}
	\begin{tabular}{c}
	\begin{minipage}{0.4\hsize} 
	\begin{center}
	\includegraphics[width=70mm,clip]{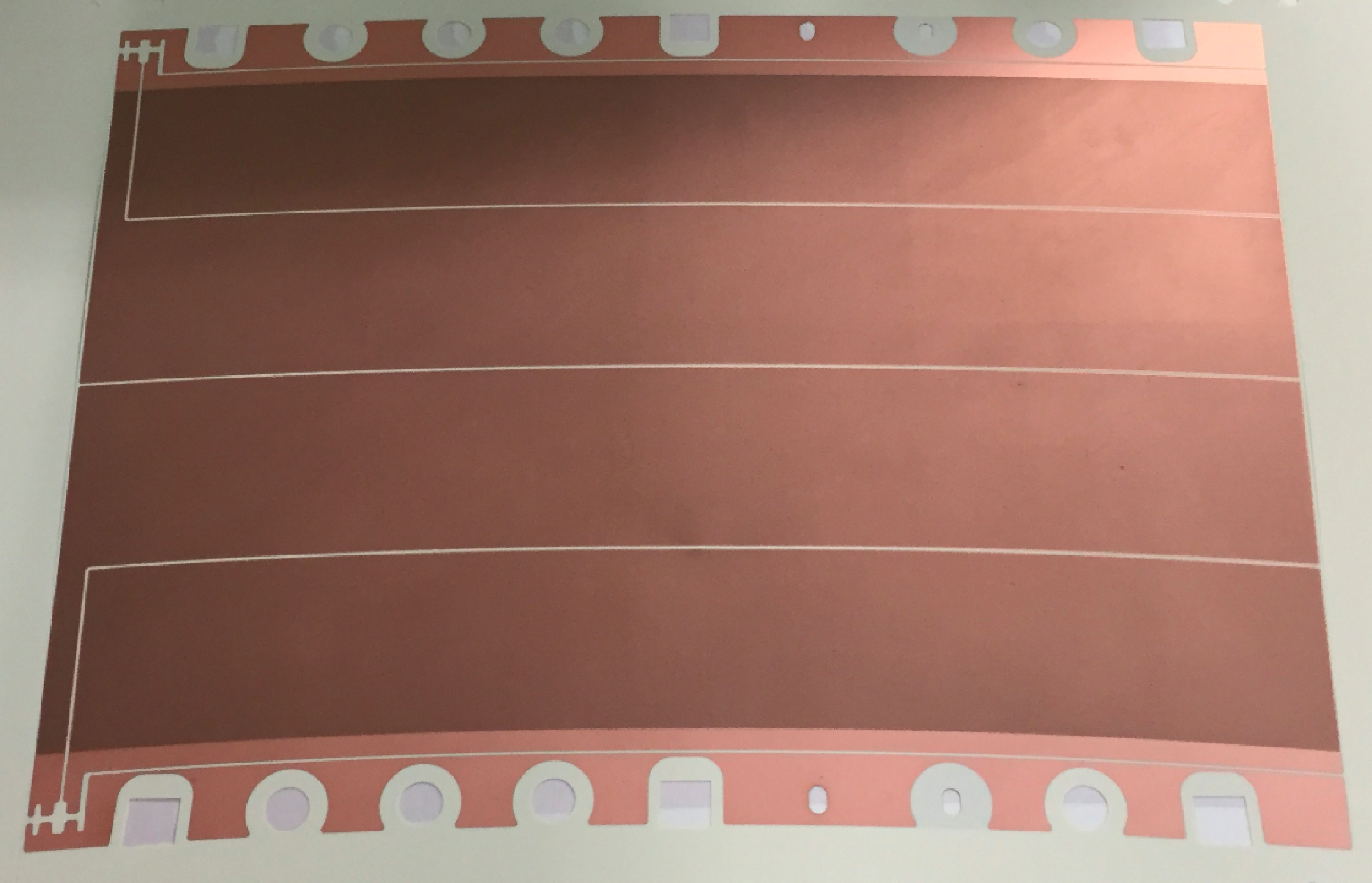}
	\end{center}
	\end{minipage}
	\begin{minipage}{0.6\hsize}  
	\begin{center}
	\includegraphics[width=80mm,clip]{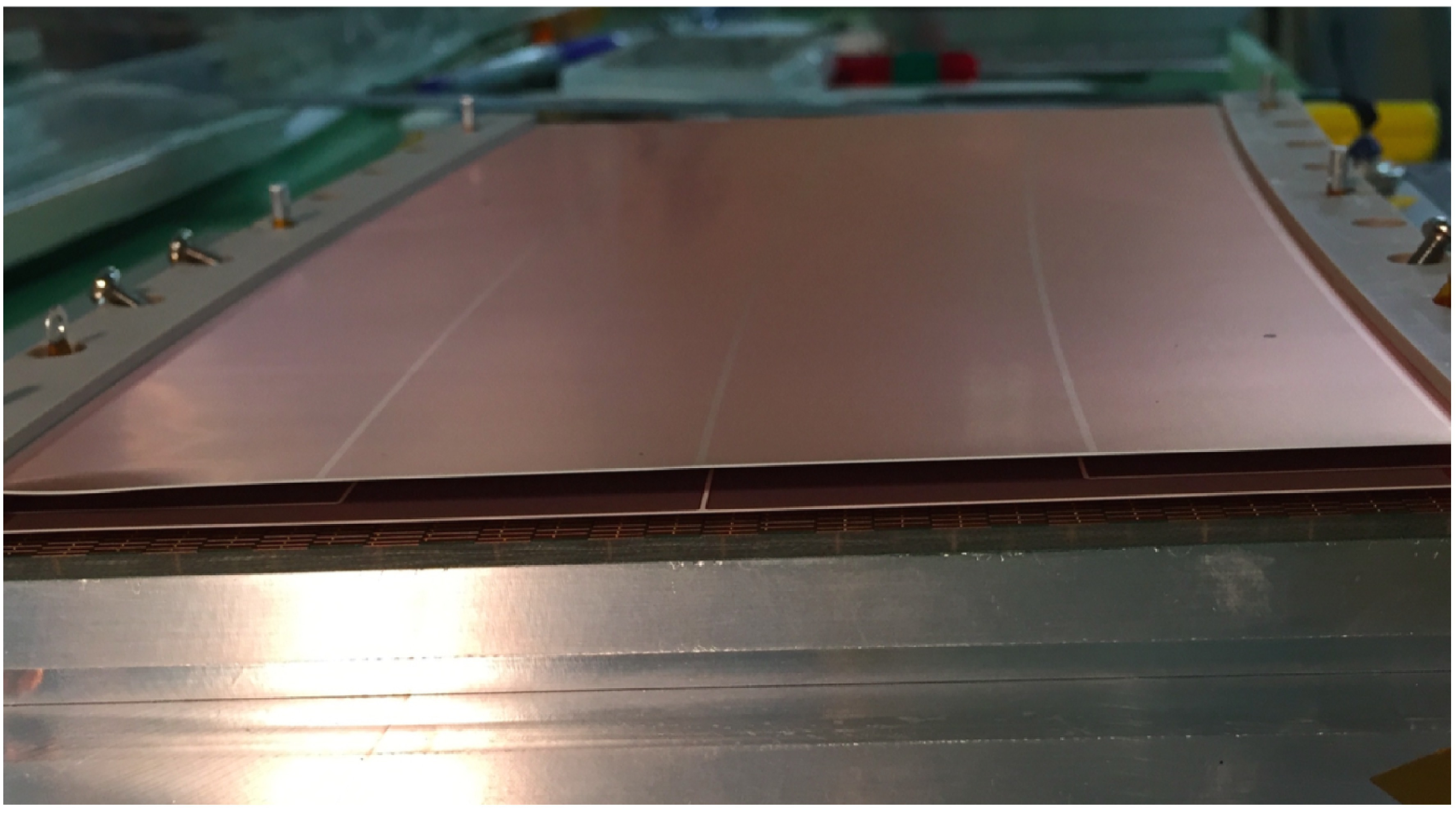}
	\end{center}
	\end{minipage}
	\end{tabular}
\end{center}
\vspace{-2mm}
\caption{The top view of the 100 $\mu m$ thick GEM and the side view of the double stacked configuration of the thick GEM. The active area of the GEM foil is about 20$\times$15 $\rm{cm^{2}}$, which is the real module size for the ILC-TPC.}
\label{fig:fig1}
\end{figure}
The design using the double thick GEM structure is considered as a candidate of an amplification device of TPC for ILC. The insulator of the 100 $\mu m$ thick GEM is made by Liquid Crystal Polymer (LCP) and the area size of one thick GEM foil is about 20$\times$15 $\rm{cm^{2}}$ like shown in the \Figref{fig:fig1}. We measured gas gain over the thick GEM foil using a $\rm{^{55}Fe}$ radiation source in order to check its gas gain uniformity. Gas gain is, off course, affected by external environment such as temperature and pressure. We prepared one reference point on each GEM foil because a relatively long time was needed to store statistics of signals from the source and measure gas gain over the GEM foil. If gas gain of the reference point varies depending on time, we multiplied this varied value of gas gain of the reference point as a correction factor. Unfortunately on the measurements of gas gain uniformity using several samples, we observed large non-uniformity and it reached more than 50\% difference as indicated in the \Figref{fig:fig2}.          
%
%
Here, there are several questions: Is this 50\% difference of gas gain variation due only to the effect of the thickness of the GEM foil? Are there any optimum geometries to restrain the difference of gas gain of the same GEM foil? How precisely do we have to manufacture the GEM foil in order to remove the effect of the thickness? To answer these questions we performed simulation using $\rm{garfield^{++}}$ \cite{Garf}. 
%
%
%
\begin{figure}[htbp]
\begin{center}
\includegraphics[width=164mm,clip]{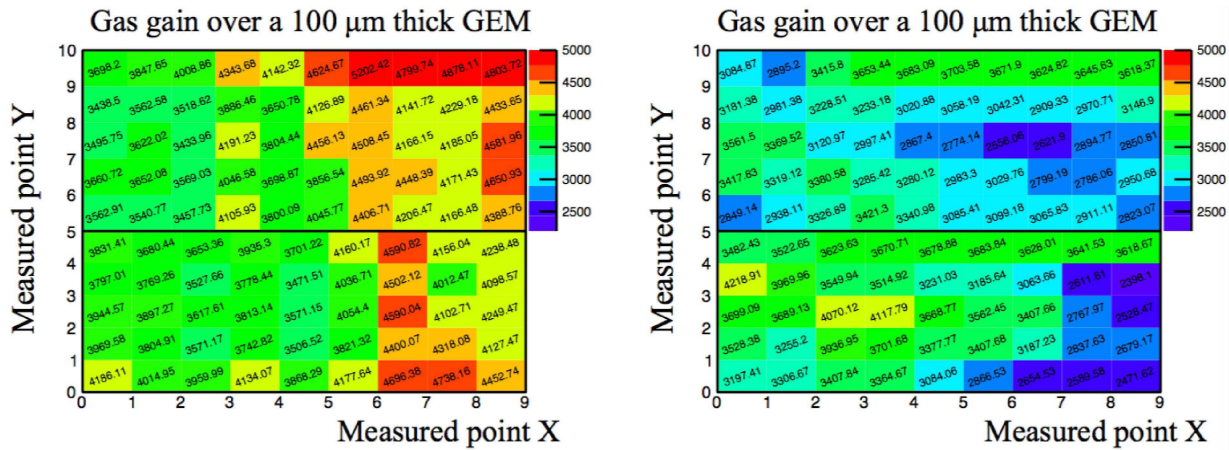}
\end{center}
\vspace{-2mm}
\caption{Gas gain uniformity of the 100 $\mu m$ thick GEM for two kinds of samples. Measurement was performed with the double stacked configuration. Gain correction depending on the measurement time is included.}
\label{fig:fig2}
\end{figure}
%
%
\section{A simulation method and geometries of the GEM} \label{Section1}
\begin{figure}[htbp]
\begin{center}
	\begin{tabular}{c}
	\begin{minipage}{0.3\hsize} 
	\begin{center}
	\includegraphics[width=52mm,clip]{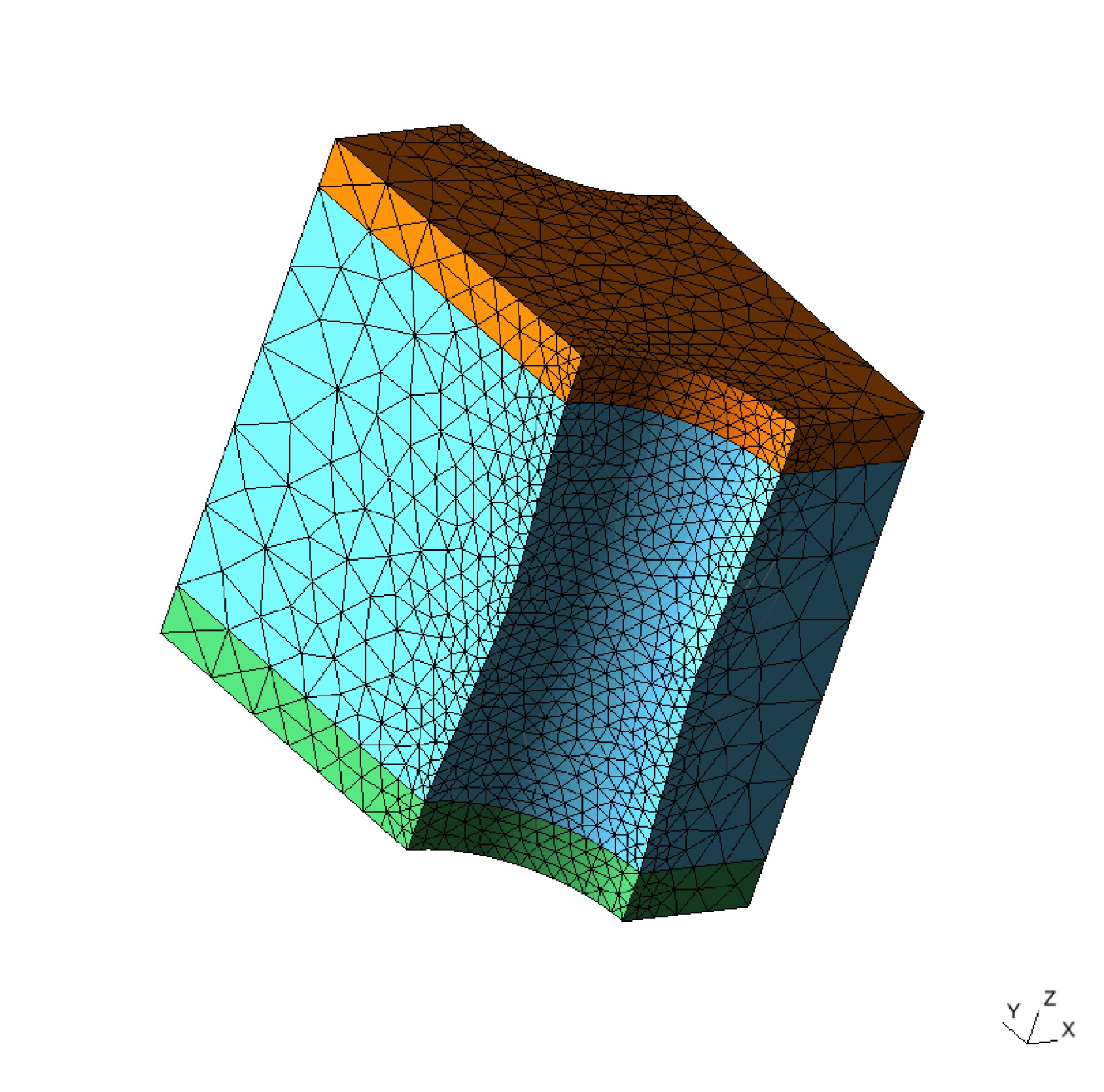}
	\end{center}
	\end{minipage}
	\begin{minipage}{0.7\hsize}  
  \begin{tabular}{ | l | c | c | } \hline
                                  & 50 $\mu m$ standard GEM& 100 $\mu m$ thick GEM \\ \hline 
Voltage [V]                 &   $ 280, 310 $      & $ 350, 380 $  \\
Dielec.  [$\mu m$]      & $44 \sim 56 $      & $90 \sim 110 $  \\
Copper [$\mu m$]        &   $ 5, 10      $     & $ 5, 10 $  \\
Hole (model)                & conical             & straight, conical  \\
Hole ($\phi$) [$\mu m$] & $60 \sim 120 $ & $70 \sim 120 $  \\ \hline
        \end{tabular}
	\end{minipage}
	\end{tabular}
\end{center}
\vspace{-2mm}
\caption{A example of the GEM geometry and the list of GEM models which are input in this simulation. Only one sector of the GEM is used for field calculation.}
\label{fig:fig3}
\end{figure}

\begin{figure}[htbp]
\begin{center}
	\begin{tabular}{c}
	\begin{minipage}{0.5\hsize} 
	\begin{center}
	\includegraphics[width=70mm,clip]{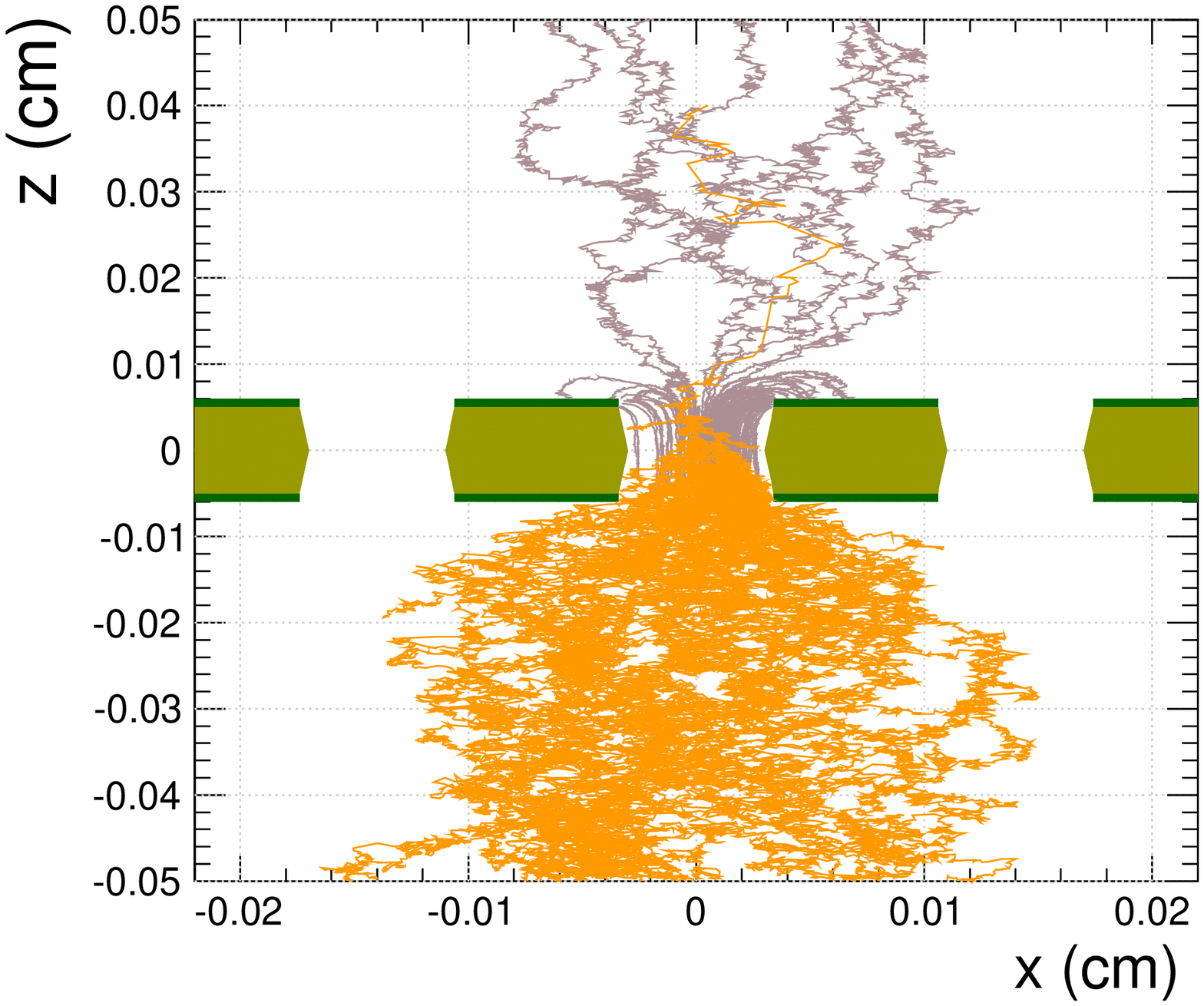}
	\end{center}
	\end{minipage}
	\begin{minipage}{0.5\hsize}  
	\begin{center}
	\includegraphics[width=70mm,clip]{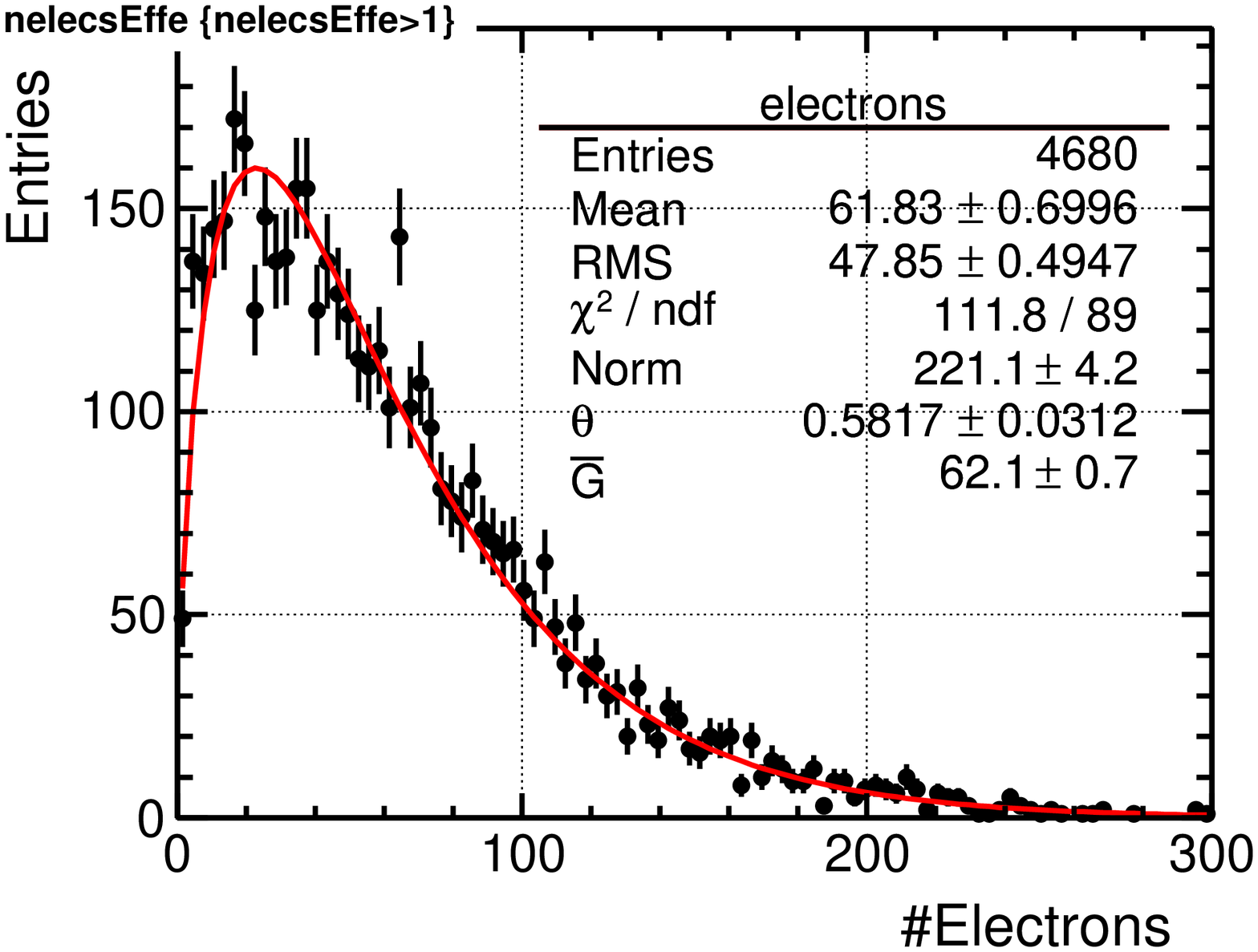}
	\end{center}
	\end{minipage}
	\end{tabular}
\end{center}
\vspace{-2mm}
\caption{The simulated avalanche amplification process where the electric field of the drift volume is set to 230V/cm. Orange lines correspond to drifting electrons and purple lines show drifting ions. Gas gain is extracted as $\bar{G}$ after fitting the Polya function.}
\label{fig:fig4}
\end{figure}

For this simulation, we made several geometries of the GEM hole using a free software called Gmsh \cite{Gmsh} and calculation of  electric fields is carried out using Elmer \cite{Elmer} after setting proper electric potentials for each geometry. Simulation of electron transportation is performed using $\rm{garfield^{++}}$. During real manufacturing of the GEM, a laser etching technique is the best way because it can shave parts around the GEM hole smoothly. The GEM holes which are used for this simulation were also made assuming a smooth conical hole etched by the laser technique. To compare the difference of gas gain which is derived from the effect of the thickness of the insulator,  the thickness of the conductor, the applied high voltages, the difference of models and diameters of the GEM hole, we prepared several geometries based on the 50 $\mu m$ standard GEM and the 100 $\mu m$ thick GEM. The geometries of each GEM model are listed in the below \Figref{fig:fig3} and the table. 
%
%
%
The gas used in this simulation is T2K TPC gas \cite{TPCgas} with NTP, which is a candidate gas for the ILC-TPC. Since T2K gas contains isobutane in the mixture, the penning effect is included also. The excitation-ionisation transfer rate should also be input into the simulation and we input a value of 0.25 as a transfer rate. We set electric fields of the drift volume 230V/cm for all geometries and electric fields of the induction volume are set 3000V/cm and 4000V/cm for the 50 $\mu m$ standard GEM and the 100 $\mu m$ thick GEM, respectively. An initial electron is inserted at Z $=$ 0.5 mm above the surface of each GEM geometry. The X and Y positions of the initial electron are set randomly in the area of each hole size of the GEM. The initial electron is sucked by the GEM hole by the electric force lines and the avalanche amplification process follows. Gas gain is estimated by fitting with the Polya function, the \Equref{eq:equ1}, after counting the produced secondary electrons which reach the volume under the GEM geometry event by event. The \Figref{fig:fig4} shows the gas gain distribution with the 100 $\mu m$ thick GEM and 350 V is applied for the GEM. The result of gas gain after fitting with the Polya function is about 62. This value is actually a bit small compared with the actual measurement using the 100 $\mu m$ thick GEM. (According to the reference \cite{Aus3}, the excitation-ionisation transfer rate was estimated for mixture gas of Ar+isobutane(5\%) and its estimated value is 0.32. If we use this value, gas gain will be almost consistent with our real measurement.)   
%
%
%
\begin{eqnarray}
f_{polya}(G; \bar{G}, \theta)  \; = \;  \frac{   (\theta+1)^{\theta+1}  }{\Gamma(\theta+1)}  \Bigl( \frac{G}{\bar{G}} \Bigr)^{\theta}    \exp\Bigl( - (\theta+1) \Bigl( \frac{G}{\bar{G}} \Bigr) \Bigr)
\label{eq:equ1}
\end{eqnarray}
%
%
\section{Impact of the thickness of the GEM geometry} \label{Section1}
\begin{figure}[h]
\begin{center}
	\begin{tabular}{c}
	\begin{minipage}{0.5\hsize} 
	\begin{center}
	\includegraphics[width=70mm,clip]{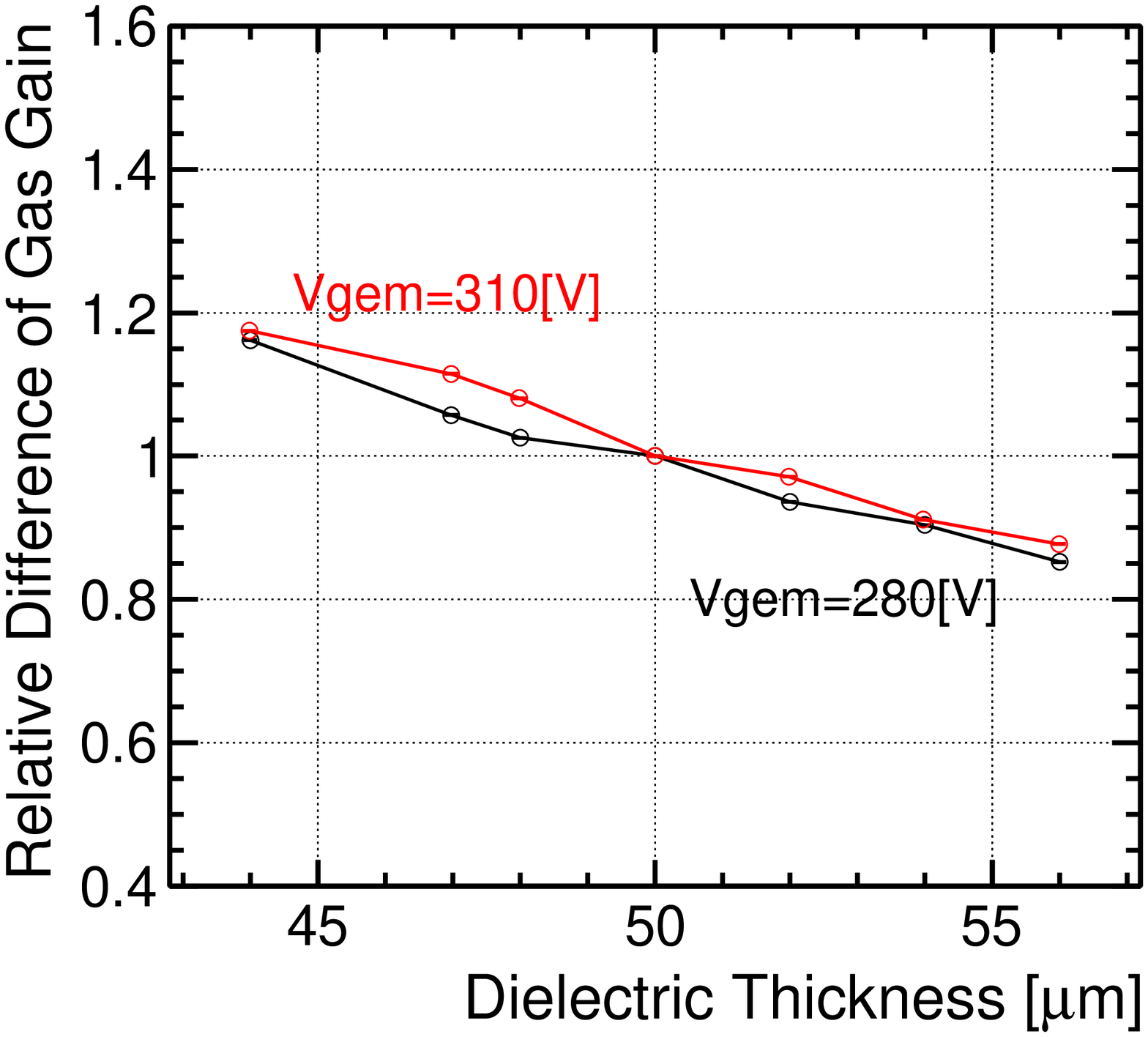}
	\end{center}
	\end{minipage}
	\begin{minipage}{0.5\hsize}  
	\begin{center}
	\includegraphics[width=70mm,clip]{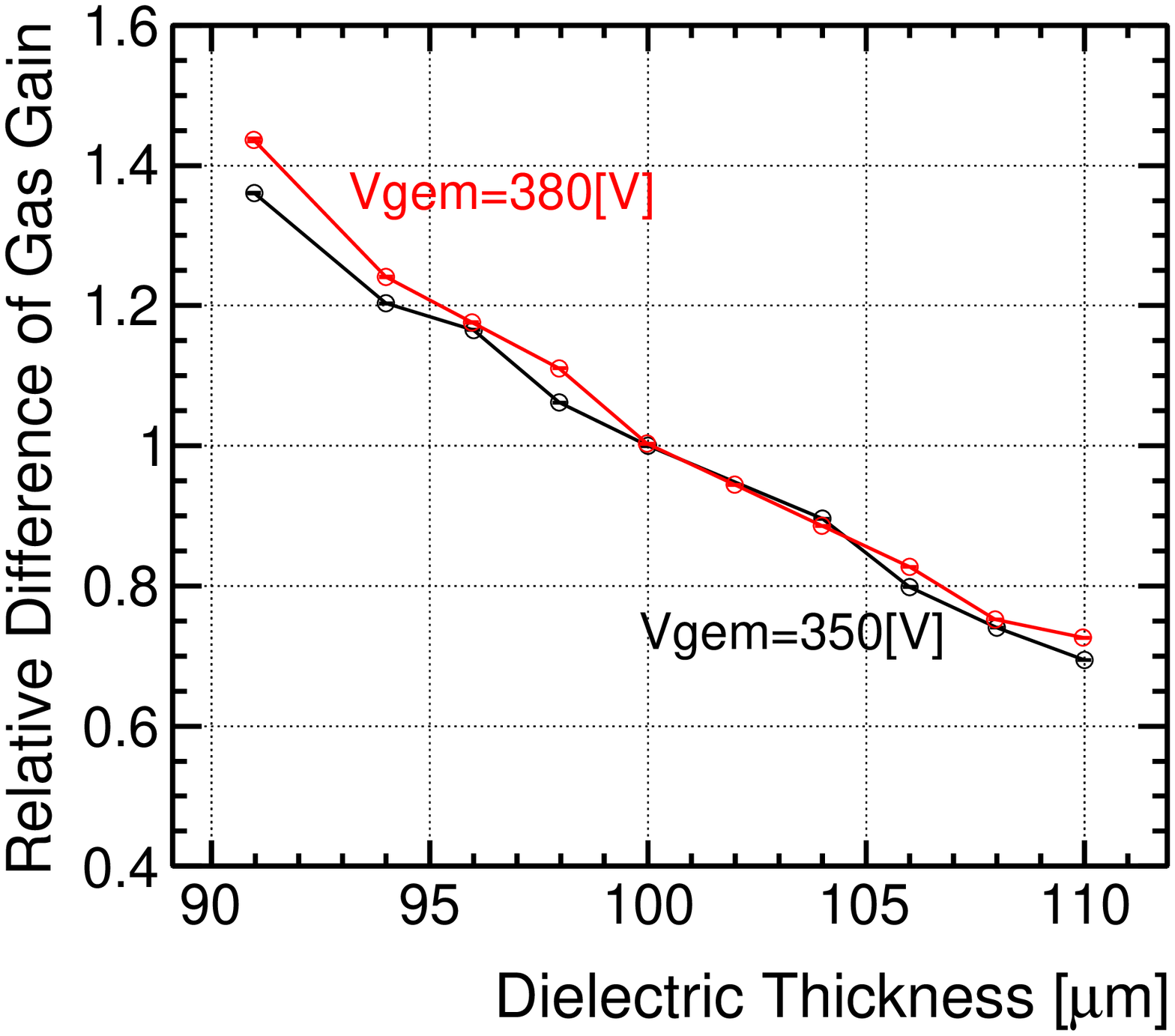}
	\end{center}
	\end{minipage}
	\end{tabular}
\end{center}
\vspace{-2mm}
\caption{The relative difference of gas gain for applied high voltages as a function of the difference of the dielectric thickness. Left and right correspond to the standard and the thick GEM, respectively.}
\label{fig:fig5}
\end{figure}

\begin{figure}[h]
\begin{center}
	\begin{tabular}{c}
	\begin{minipage}{0.5\hsize} 
	\begin{center}
	\includegraphics[width=70mm,clip]{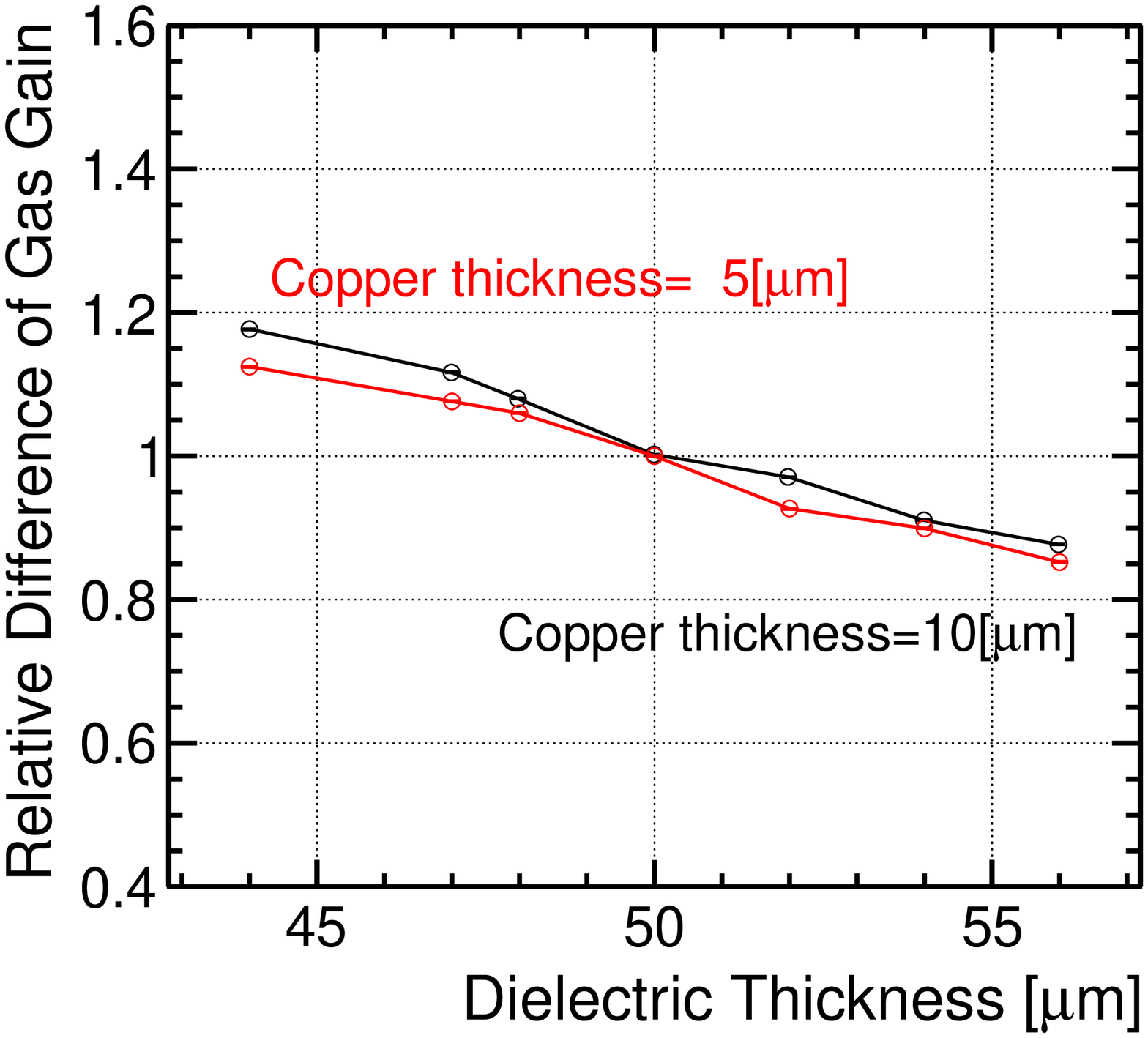}
	\end{center}
	\end{minipage}
	\begin{minipage}{0.5\hsize}  
	\begin{center}
	\includegraphics[width=70mm,clip]{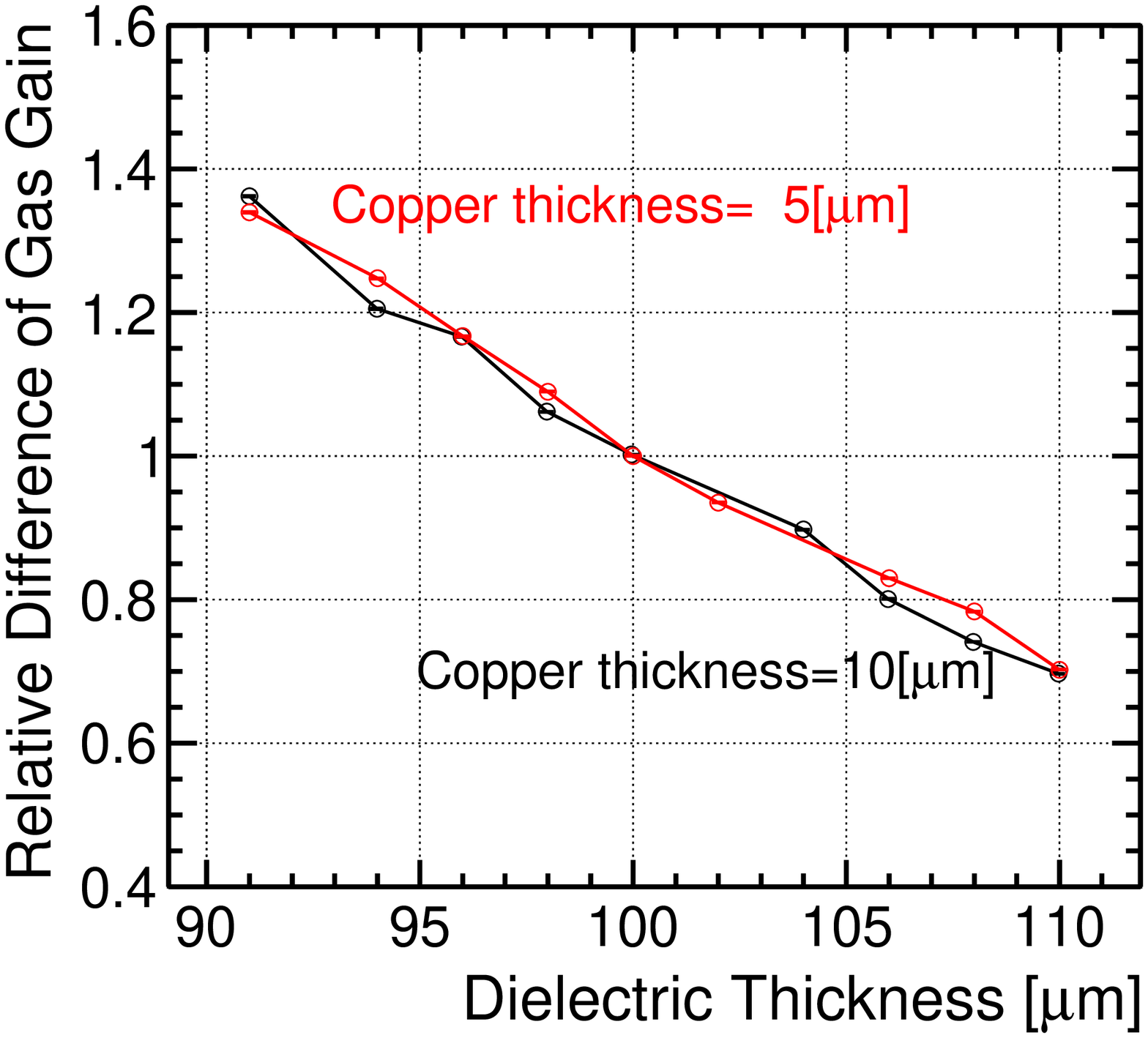}
	\end{center}
	\end{minipage}
	\end{tabular}
\end{center}
\vspace{-2mm}
\caption{The relative difference of gas gain for the thickness of the copper as a function of the difference of the dielectric thickness. Left and right show the standard and the thick GEM.}
\label{fig:fig6}
\end{figure}

\begin{figure}[htbp]
\begin{center}
	\begin{tabular}{c}
	\begin{minipage}{0.5\hsize} 
	\begin{flushleft}
	\includegraphics[width=70mm,clip]{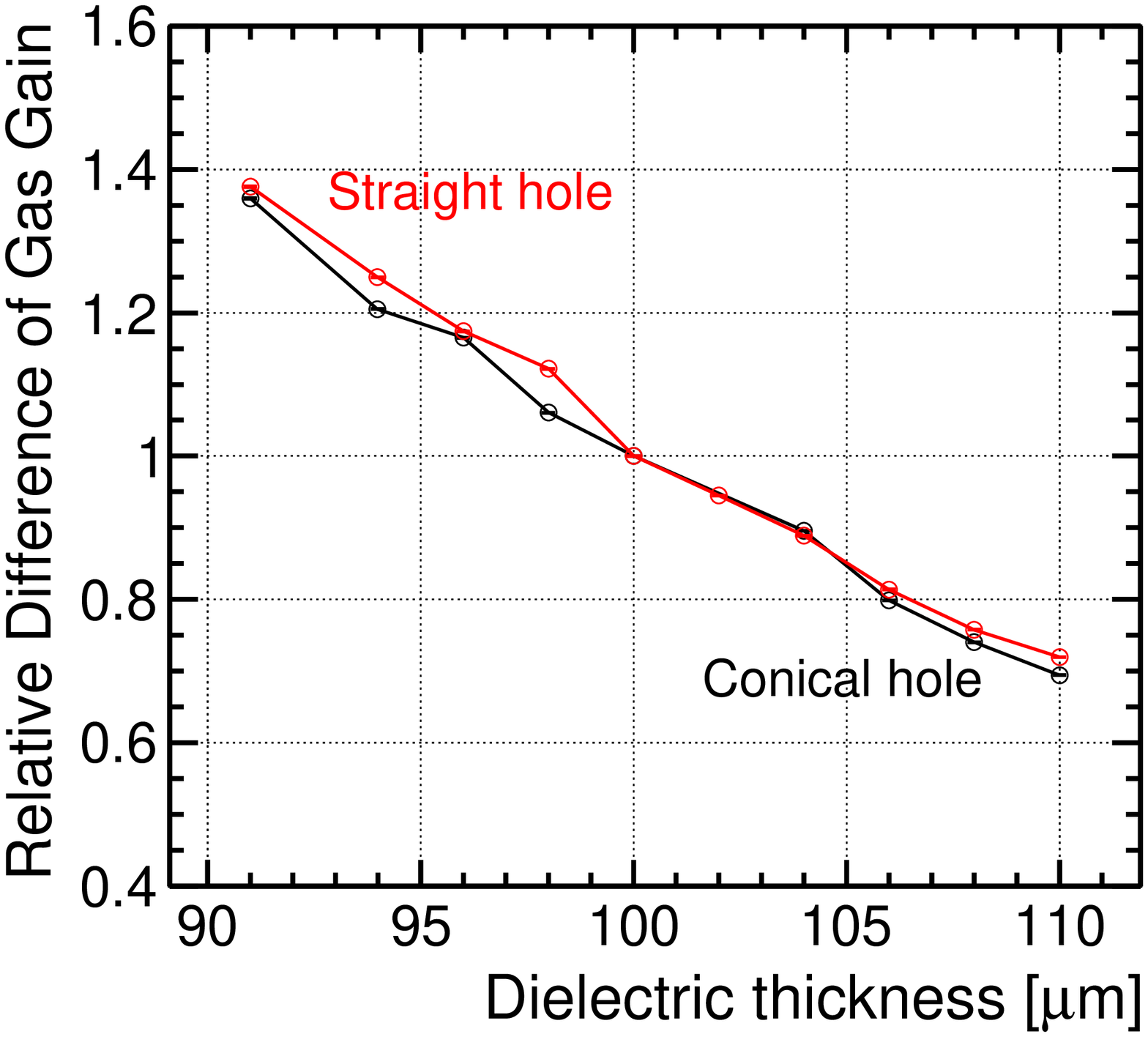}
	\end{flushleft}
\vspace{-2mm}
\caption{The relative difference of gas gain for the model of the GEM hole as a function of the difference of the dielectric thickness. Here only the result for the 100 $\mu m$ thick GEM is shown.}
\label{fig:fig7}
	\end{minipage}
	\begin{minipage}{0.5\hsize}  
	\begin{center}
	\includegraphics[width=70mm,clip]{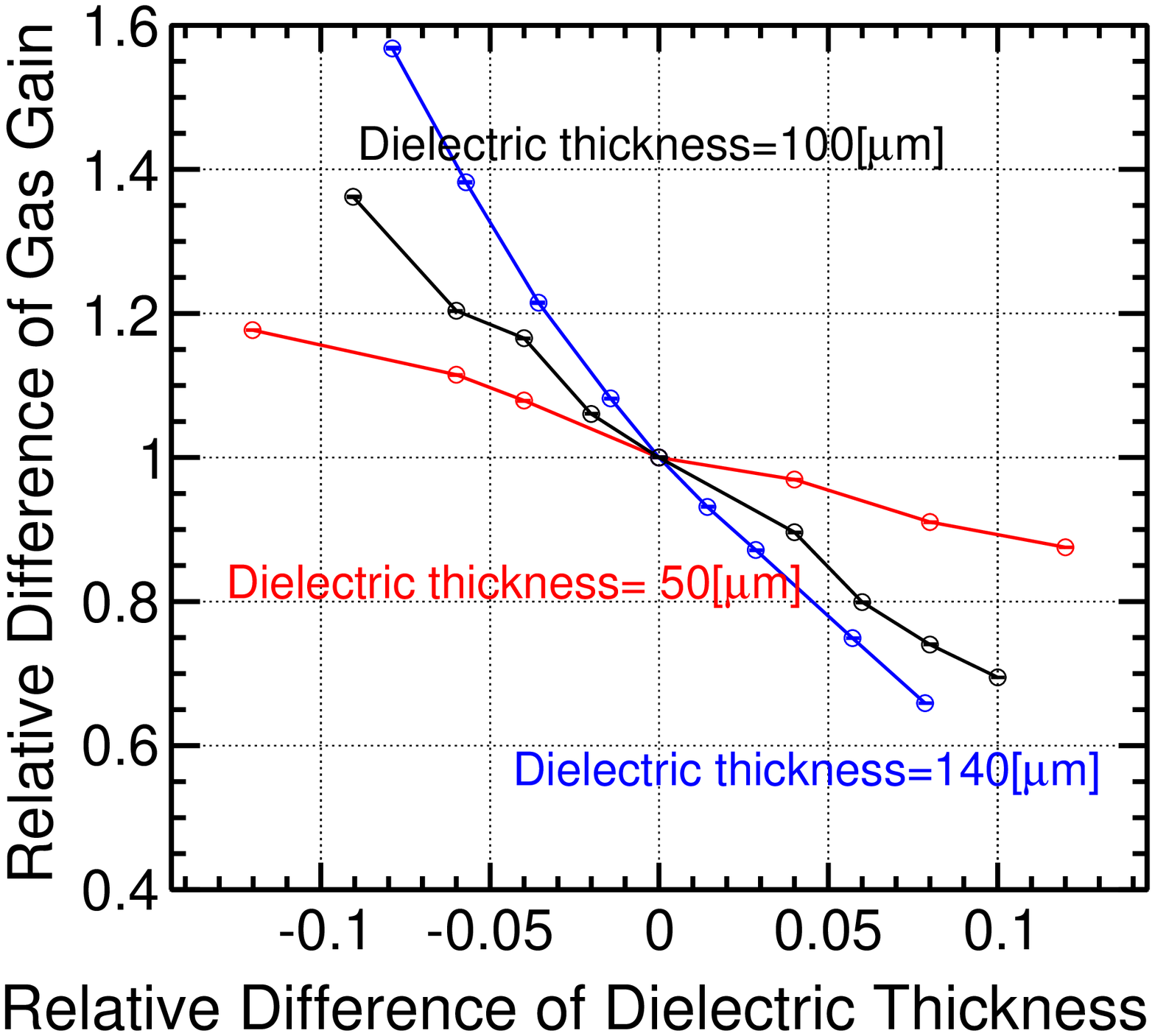}
	\end{center}
\vspace{-2mm}
\caption{The relative difference of gas gain as a function of the relative difference of the dielectric thickness for the 50 $\mu m$, 100 $\mu m$, and 140 $\mu m$ thick GEMs. }
\label{fig:fig8}
	\end{minipage}
	\end{tabular}
\end{center}
\end{figure}

First of all we tried to understand the effect of the thickness of the insulator for applied high voltages using the models of the 50 $\mu m$ standard GEM and the 100 $\mu m$ thick GEM. The \Figref{fig:fig5} show the relative difference of gas gain when the thickness of the insulator changed. It is seen that the relative difference of gas gain does not change depending on the applied high voltages for the both standard GEM and thick GEM. The relative difference of gas gain in which the thickness of the conductor changed are also same for the both of the GEMs as illustrated by the \Figref{fig:fig6}. Similarly the ratio of the difference of gas gain does not depend on the model of the GEM hole, which is shown in the \Figref{fig:fig7}. But a remarkable thing is that a degree of slopes along the thickness for the both of GEMs are different. In the case of the 50 $\mu m$ standard GEM, the maximum plateau area of the high electric field along the GEM hole where avalanche amplification become possible is originally narrow. Even if the thickness of the insulator is a bit different in some places, it seems that it does not cause the large difference of gas gain. In the case of the 100 $\mu m$ thick GEM, since the maximum plateau area is wider originally, the change of the area of the high electric field which is derived from the change of the thickness of the insulator has greater effect on a degree of avalanche amplification. Eventually the difference of avalanche amplification on the 100 $\mu m$ thick GEM is larger than that of the 50 $\mu m$ standard GEM.
%
%
%
\\
\\
As already mentioned above, it was understood that the difference of gas gain strongly depends on the thickness of the insulator and it is not affected by the difference of the other geometrical parameters. Although there are no official reports on manufacturing accuracy of the thickness of the insulator that LSP is used. According to the study \cite{Aus4}, the accuracy of the thickness of a small size 8$\times$3 $\rm{cm^{2}}$ 100 $\mu m$ thick GEM whose insulator is LSP is 100$\pm$5 $\mu m$. If there is 5 \% difference on the thickness of the 100 $\mu m$ thick GEM, 20 \% difference of gas gain is potentially anticipated only taking the effect of the thickness of the GEM foil into account. The size of the GEM foils which are shown in the introduction of this paper is 20$\times$15 $\rm{cm^{2}}$. The more than 50 \% difference of gas gain in the same GEM foil can be possible when we consider the two facts that the size of the GEM foil is larger than that of the reference \cite{Aus4} and the double stacked configuration is used. 
%
%
In contrast, assuming that there is the difference of the thickness for the both standard and 100 $\mu m$ thick GEM at the same level, the standard GEM can prevent the difference of gas gain more than two times better compared with the 100 $\mu m$ thick GEM. According to spec information of the standard GEM which is produced by CERN \cite{Aus5}, because accuracy of manufacturing of the 50 $\mu m$ thick insulator whose material is Polyimide reach 50$\pm$1 $\mu m$ which is about 2 \%, the intrinsic variation of gas gain will be about 2 \%. If we want to get the same degree of gas gain uniformity with the 50 $\mu m$ standard GEM using the 100 $\mu m$ thick GEM, the \Figref{fig:fig8} indicates that it is necessary to manufacture the thickness of the insulator with accuracy of less than 1 \%. 
%
\newpage
\section{Impact of the hole diameter of the GEM} \label{Section1}
\begin{figure}[ht]
\begin{center}
	\begin{tabular}{c}
	\begin{minipage}{0.5\hsize} 
	\begin{center}
	\includegraphics[width=76mm,clip]{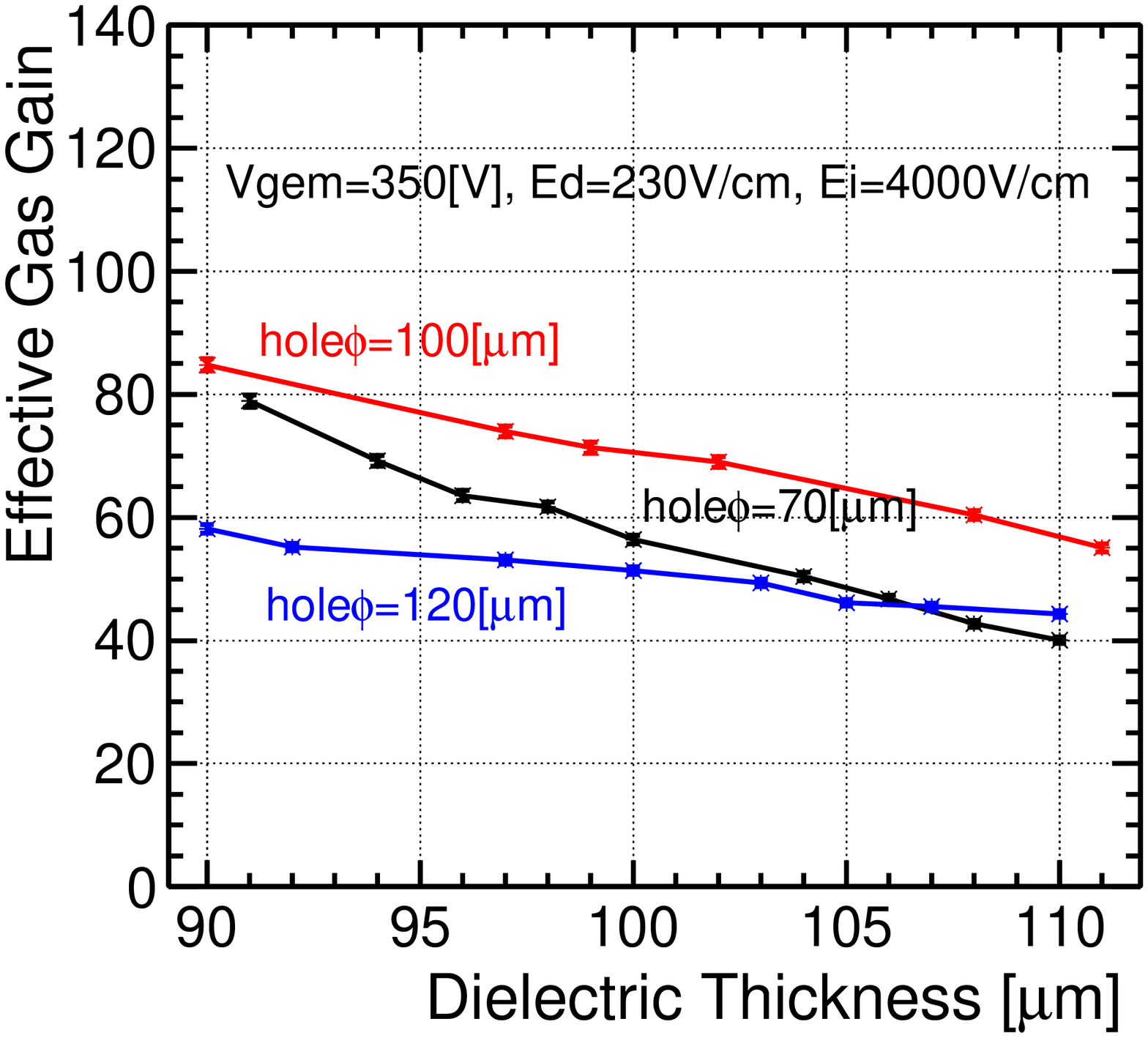}
	\end{center}
	\end{minipage}
	\begin{minipage}{0.5\hsize}  
	\begin{center}
	\includegraphics[width=76mm,clip]{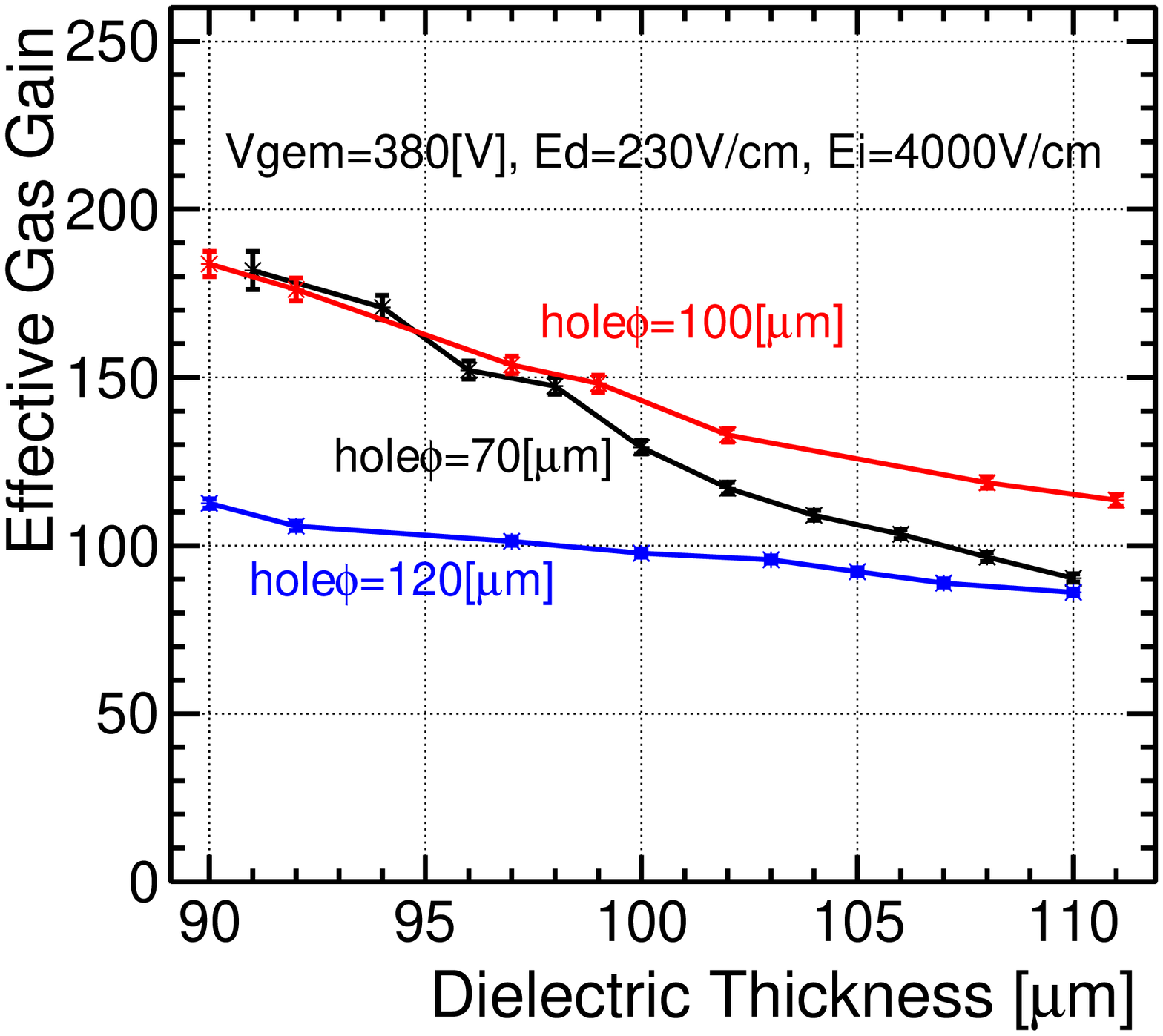}
	\end{center}
	\end{minipage}
	\end{tabular}
\end{center}
\vspace{-2mm}
\caption{Effective gas gain as a function of the dielectric thickness for the different diameters of the GEM hole. The applied high voltages to the GEM are 350V and 380V for the left and right figures.}
\label{fig:fig9}
\end{figure}
\begin{figure}[ht]
\begin{center}
	\begin{tabular}{c}
	\begin{minipage}{0.5\hsize} 
	\begin{center}
	\includegraphics[width=70mm,clip]{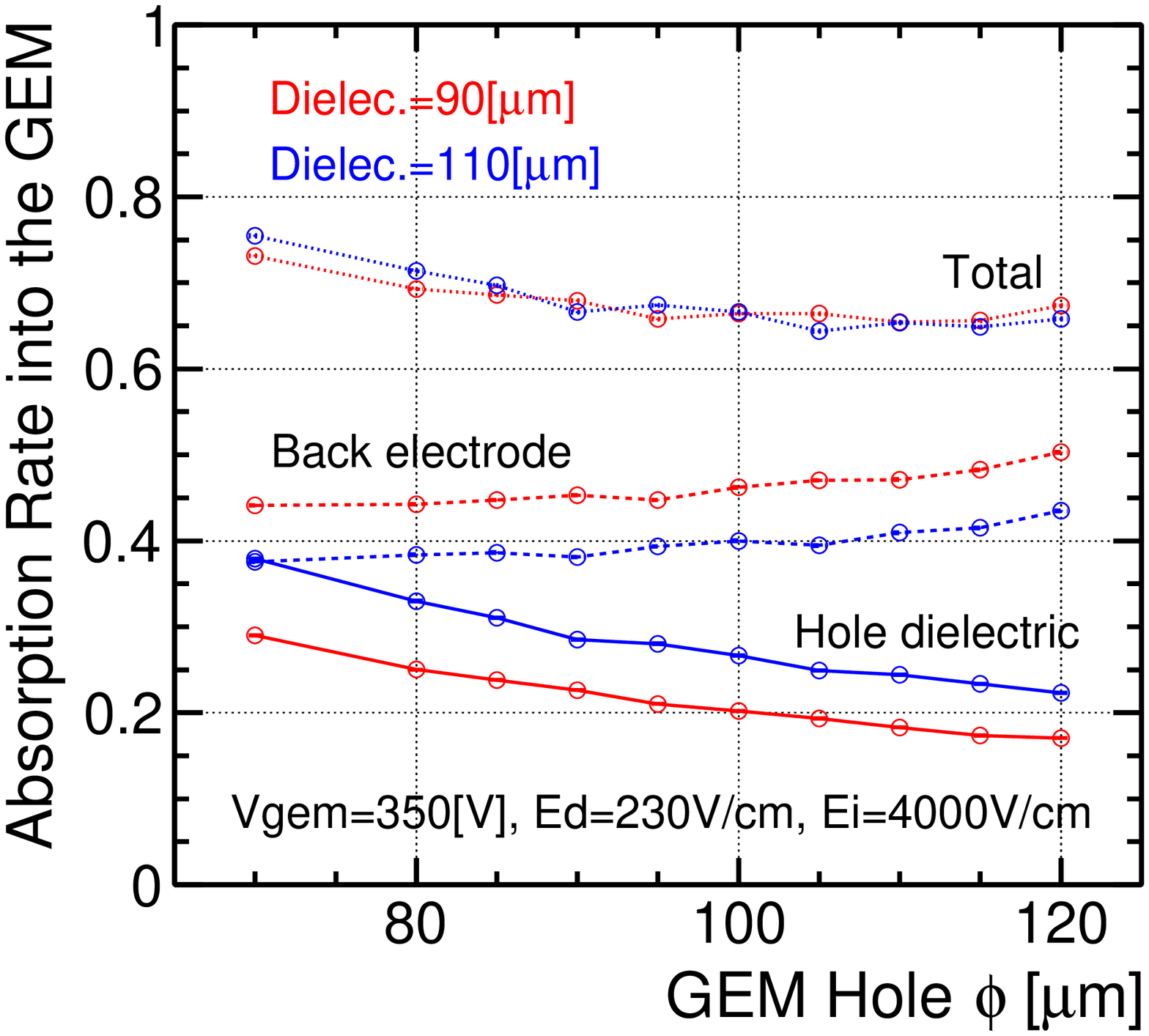}
	\end{center}
	\end{minipage}
	\begin{minipage}{0.5\hsize}  
	\begin{center}
	\includegraphics[width=75mm,clip]{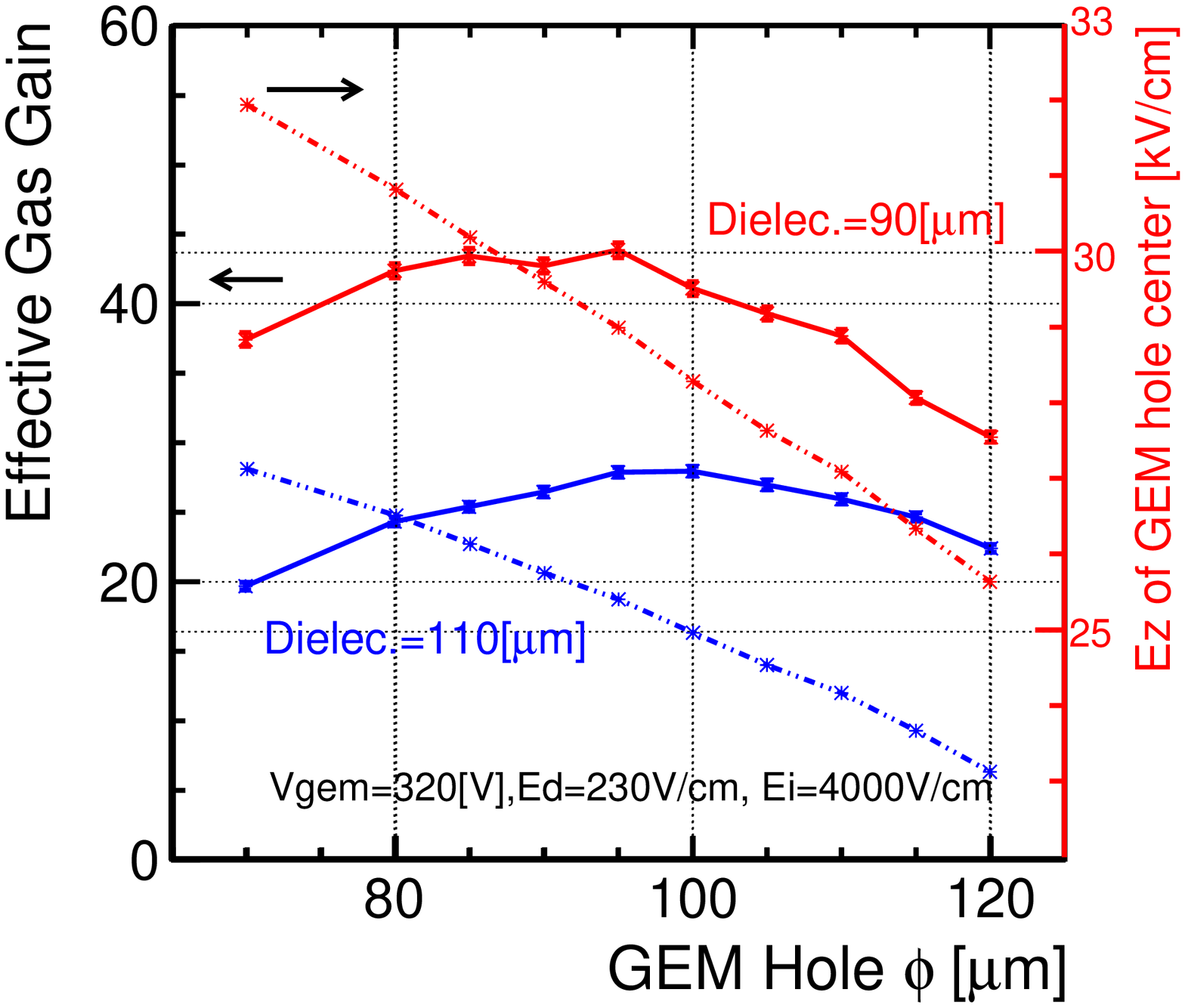}
	\end{center}
	\end{minipage}
	\\
	\begin{minipage}{0.5\hsize}  
	\begin{center}
	\includegraphics[width=70mm,clip]{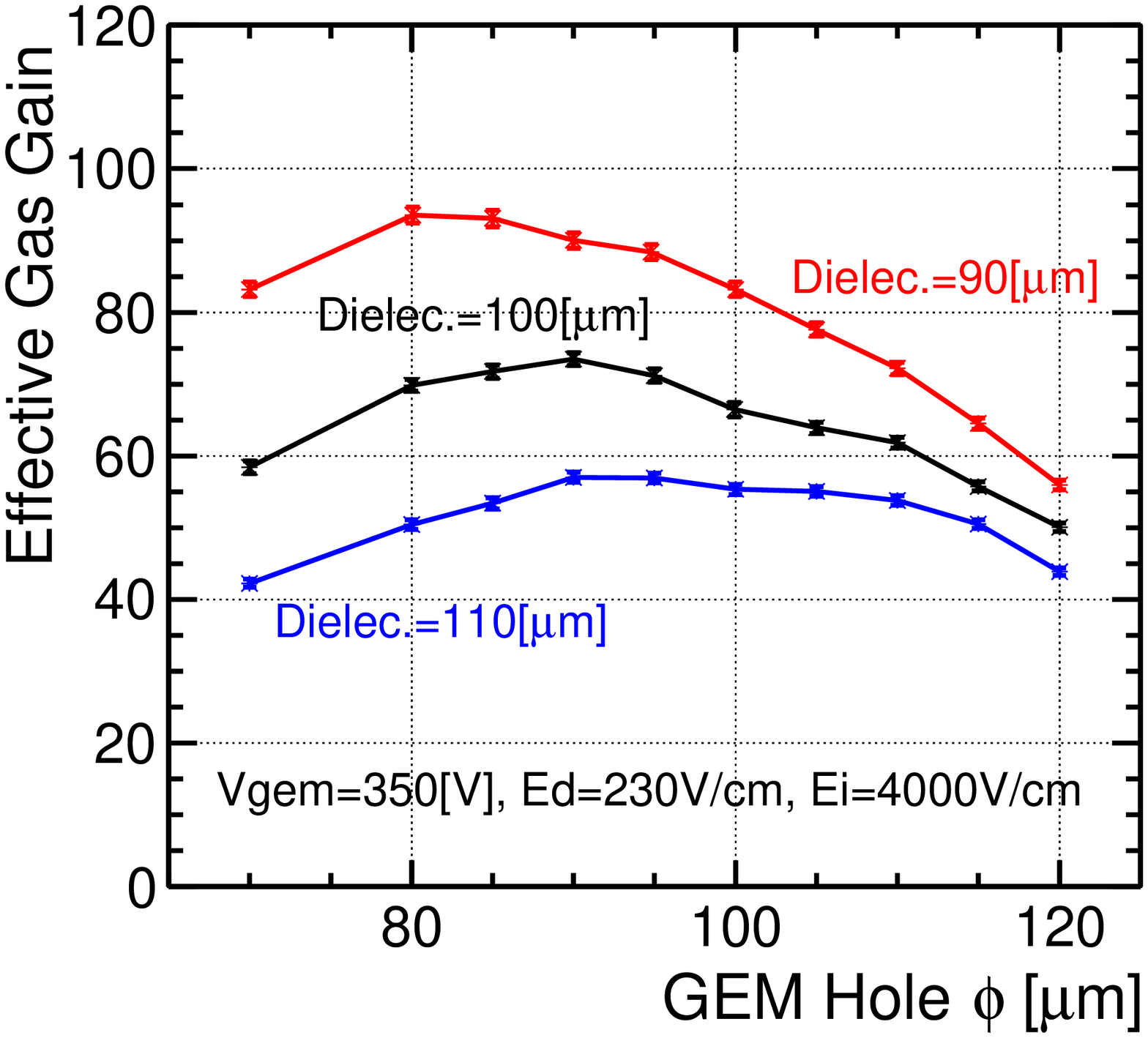}
	\end{center}
	\end{minipage}
	\begin{minipage}{0.5\hsize}  
	\begin{center}
	\includegraphics[width=70mm,clip]{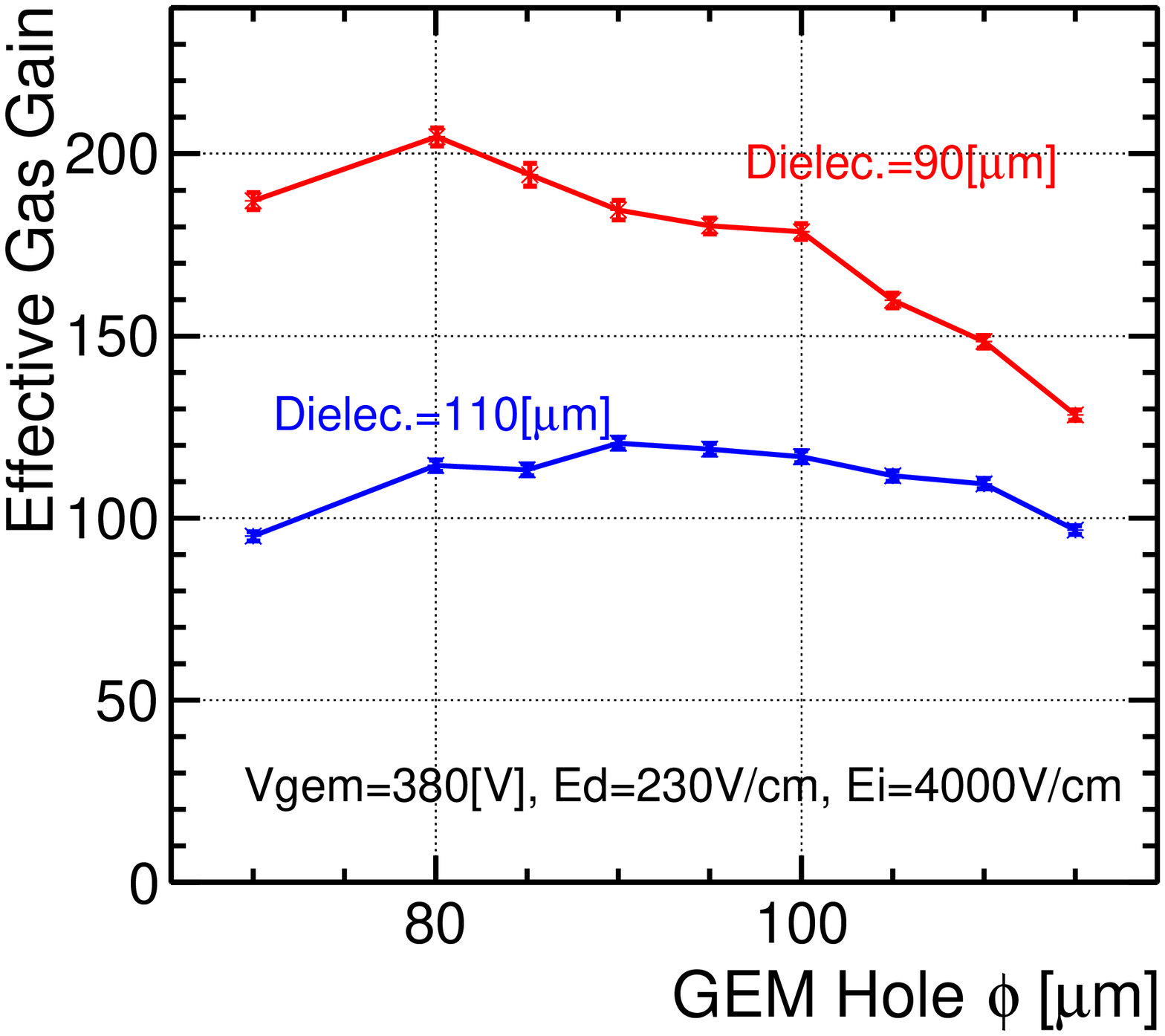}
	\end{center}
	\end{minipage}
	\end{tabular}
	\vspace{-2mm}
	\caption{The top left plot shows the absorption rate of produced secondary electrons onto the back electrode and the dielectric surface of the GEM. The top right, bottom left and right plots are effective gas gain as a function of the diameter of the GEM hole for the different dielectric thickness. The three plots show the variation as a function of gas gain. The GEM diameter of 90$\sim$100 $\mu m$ gives a stable value of gas gain for different of the dialectic thickness. The similar tendency is appeared for each gas gain. }
\label{fig:fig10}
\end{center}
\end{figure}

We also investigated the difference of gas gain between the thickness of the insulator and the size of the GEM hole. The clear difference of gas gain was observed when the size of the GEM hole changed. As shown in the \Figref{fig:fig9}, in the case that the thickness of the insulator become thicker than the base line, the gas gain is recovered when the size of the GEM hole is increased up to 100 $\mu m$. Even if applied high voltages to the GEM are changed, the same results are obtained. This can be understood if the absorption of secondary electrons into the side surface or bottom surface of the GEM hole decreased because the size of the GEM hole was enlarged. Such tendency is appeared in the top left plot on the  \Figref{fig:fig10}.  The definition of an absorption rate on the plot is simply $R= N_{abs}/N_{aval}$, where $N_{abs}$ means the number of electrons which reach to the back electrode or the dielectric surface of the GEM and $N_{aval}$ corresponds to the number of produced electrons by the avalanche process. The absorption rate of produced electrons into the dielectric surface of the GEM gradually decreases up to the diameter of 90 $\mu m$. The simulated difference of the rate  is about 10\% compared to the standard diameter of 70 $\mu m$. 
%
%
\begin{figure}[htbp]
\begin{center}
	\begin{tabular}{c}
	\begin{minipage}{0.5\hsize} 
	\begin{center}
	\includegraphics[width=70mm,clip]{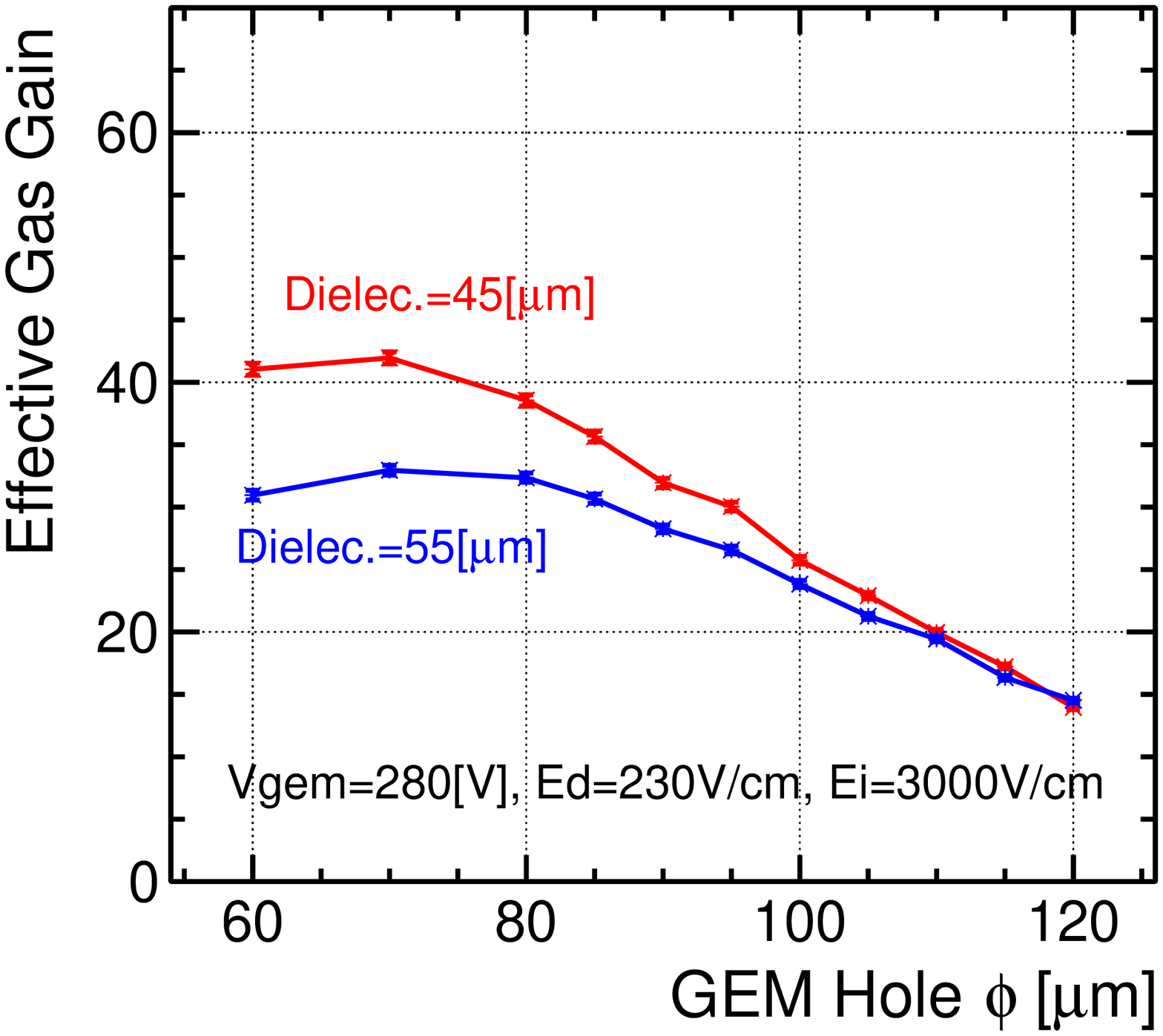}
	\end{center}
	\end{minipage}
	\begin{minipage}{0.5\hsize}  
	\begin{center}
	\includegraphics[width=70mm,clip]{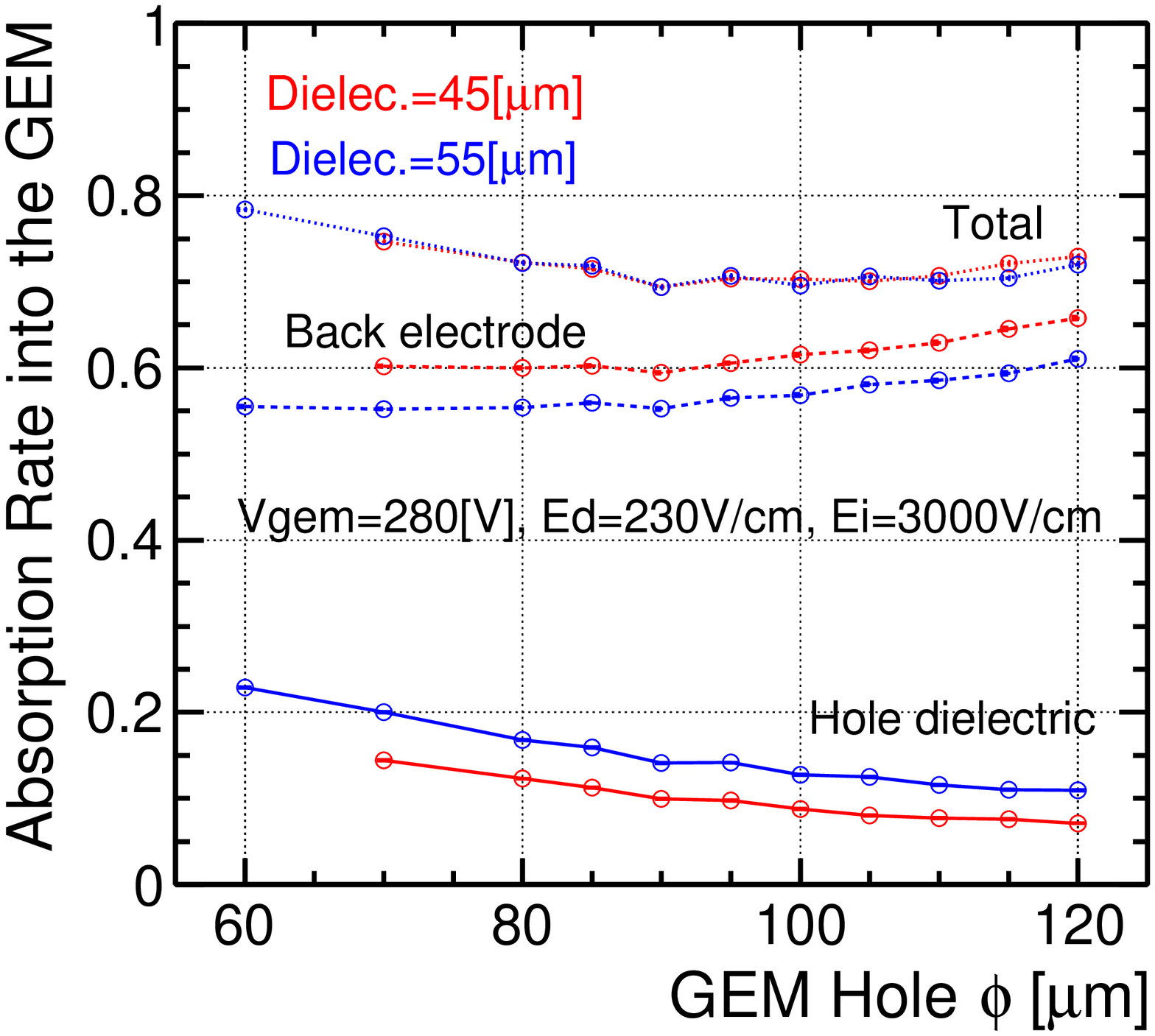}
	\end{center}
	\end{minipage}
	\end{tabular}
\end{center}
\vspace{-2mm}
\caption{Effective gas gain and the absorption rate of produced secondary electrons as a function of the diameter of the GEM hole for the different dielectric thickness are shown on the left and right plots.}
\label{fig:fig11}
\end{figure}
\newpage
The top right, bottom left and right pots on the \Figref{fig:fig10} show effective gas gain as a function of the diameter of the GEM hole for the different dielectric thicknesses. We simulated three patterns which are low, middle and high gain. In any case gas gain of each dielectric thickness gradually increase and reach plateau around the diameter of 90 $\mu m$ where the absorption rate of electrons onto the GEM surface ceases to decrease, after which the gas gain starts to drop off. This is probably because electric fields near the center area of the GEM hole is too weak to generate much avalanche processes. The \Figref{fig:fig11} shows that the value of 70 $\mu m$ for the diameter of the 50 $\mu m$ standard GEM is almost near an optimal value in terms of the effect of the thickness of the insulator and the size of the GEM hole. In the past real measurement using the 50 $\mu m$ standard GEM with several diameters, 70 $\mu m$ was confirmed to be the most optimal diameter for the standard GEM in terms of effective and absolute gas gain \cite{Aus6}. 
\section{Conclusions} \label{Section1}
\begin{figure}[htbp]
	\begin{center}
	\includegraphics[width=78mm,clip]{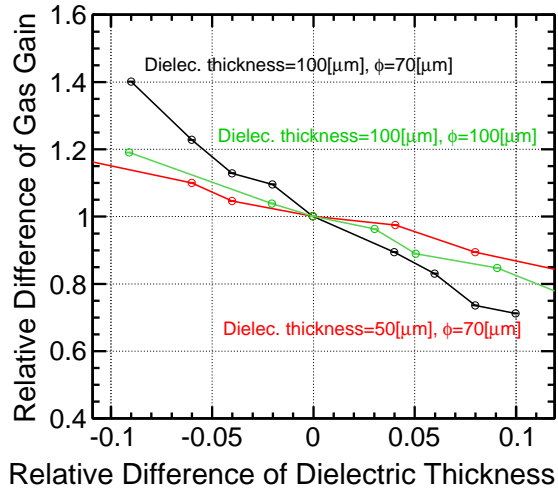}
	\end{center}
\vspace{-2mm}
\caption{The relative difference of gas gain as a function of the relative difference of the dielectric thickness for the 50 $\mu m$ and 100 $\mu m$ thick GEMs with the hole diameters of 70 $\mu m$ and 100 $\mu m$.}
\label{fig:fig12}
\end{figure}

If we want to reduce the effect of the difference of gas gain which is derived from the difference of the thickness of the insulator for the 100 $\mu m$ thick GEM, the \Figref{fig:fig10} indicates that it is probably needed to set the diameter of the GEM hole to about 90$\sim$100 $\mu m$. However, if we want to get the same degree of variation of gas gain, using the 100um thick GEM, as with the standard 50um GEM, it is necessary to manufacture the thickness of the insulator of the GEM with an accuracy of about 1$\sim$2 \% as shown in the \Figref{fig:fig12}.
%
%






\end{document}